\RequirePackage{lineno}
\documentclass[twocolumn,nofootinbib,floatfix,superscriptaddress]{revtex4-1}

\usepackage{amsmath,amssymb}
\usepackage{cancel}
\usepackage{graphicx}
\usepackage{epsfig}
\usepackage{amsmath,amsfonts,amssymb}
\usepackage{multirow}
\usepackage{colortbl}
\usepackage{hhline}
\usepackage{booktabs}
\usepackage{bbm}
\usepackage{bm}
\usepackage{tabulary}
\usepackage{pifont}
\usepackage{bbold}
\usepackage{pifont}

\newcommand{\MR}{\mathbf{M}_R}
\newcommand{\Mell}{\mathbf{M}_\ell}
\newcommand{\Ynu}{\mathbf{Y}_\nu}
\newcommand{\Ynut}{\tilde{\mathbf{Y}}_\nu}
\newcommand{\Yl}{\mathbf{Y}_\ell}

\newcommand{\Mnu}{\mathbf{M}_\nu}

\newcommand{\Unu}{\mathbf{U}_\nu}
\newcommand{\Ul}{\mathbf{U}_\ell}
\newcommand{\U}{\mathbf{U}}
\newcommand{\V}{\mathbf{V}_\delta}

\newcommand{\dmatm}{\Delta m^2_{31}}
\newcommand{\dmsol}{\Delta m^2_{21}}

\newcolumntype{K}[1]{>{\centering\arraybackslash}m{#1}}
\begin{document}

\title{New approach to neutrino masses and leptogenesis with Occam's razor}

\author{D. M. Barreiros}
\affiliation{Departamento de F\'{\i}sica and CFTP, Instituto Superior T\'ecnico, Universidade de Lisboa, Lisboa, Portugal}
\author{F. R. Joaquim}
\affiliation{Departamento de F\'{\i}sica and CFTP, Instituto Superior T\'ecnico, Universidade de Lisboa, Lisboa, Portugal}
\author{T. T. Yanagida}
\affiliation{Tsung-Dao Lee 
Institute, Shanghai Jiao Tong University, Shanghai 200240, China}
\affiliation{Kavli IPMU (WPI), UTIAS, The University of Tokyo, 5-1-5 Kashiwanoha, Kashiwa, Chiba 277-8583, Japan}


\begin{abstract}
The minimal type-I seesaw framework with texture-zero Yukawa and mass matrices inspired by Occam's razor is incompatible with normally-ordered neutrino masses (currently preferred by data) when lepton mixing originates solely from the neutrino sector. Moreover, the lightest right-handed neutrino mass required to generate the observed baryon asymmetry of the Universe via leptogenesis ($M_1 \sim 10^{14}$~GeV) is in conflict with vanilla scenarios for (Peccei-Quinn) axion dark matter where the reheating temperature of the Universe is typically below $10^{12}$~GeV. In this work, we present a new Occam's razor setup which overcomes these problems by including charged-lepton mixing parametrized by a single angle, which is predicted to be very close to the quark Cabibbo angle. Furthermore, the atmospheric mixing angle lies in the second octant and the leptogenesis scale is lowered to $\sim 5.5\times 10^{10}$~GeV, lifting the tension with the axion dark-matter hypothesis.
\end{abstract}

\maketitle

The discovery of neutrino oscillations, and the ensuing fact that neutrinos are massive particles, provide irrefutable evidence for physics beyond the Standard Model (SM). Besides being unable to accommodate neutrino masses and lepton mixing, the SM also fails at supplying a suitable dark-matter (DM) candidate, and a convincing explanation for the observed excess of matter over antimatter in the Universe. Remarkably, two of the said limitations may be overcome adding to the SM two heavy right-handed (RH) neutrinos which, besides acting as light-neutrino mass mediators at the classical level (seesaw mechanism~\cite{seesaw}), also play a crucial role in generating the baryon asymmetry of the Universe (BAU) via leptogenesis~\cite{Fukugita:1986hr}. This minimal setup provides a perfect template to address the neutrino mass and BAU problems in a rather economical and minimal way~\cite{Frampton:2002qc}.

Over the last few decades, neutrino oscillation experiments have been collecting data, gradually improving the measurement of neutrino parameters. Still, and in spite of the exceptional results obtained so far, three crucial aspects remain to be clarified: the type of neutrino mass ordering (normal or inverted), the value of the leptonic CP-violating phase $\delta$, and the $\theta_{23}$ (atmospheric mixing angle) octant. At present, global analyses of the data~\cite{Capozzi:2020qhw,deSalas:2020pgw,Esteban:2020cvm} indicate a preference for a normally-ordered (NO) neutrino mass spectrum at the $3\sigma$ level, and also for $\theta_{23}$ belonging to the second octant. Furthermore, the current best-fit value for $\delta$ lies around $3\pi/2$ for inverted-ordered (IO) neutrino mass spectrum, hinting to large CP-violating effects in the lepton sector for that case. For NO, global fits using the latest data presented at the Neutrino 2020 conference~\cite{Acero:2019ksn,NovaTalk,Abe:2019vii,T2KTalk} reveal that $\delta$ is closer to the CP conserving value $\delta\simeq \pi$ due to a mild tension between T2K and NO$\nu$A data~\cite{Capozzi:2020qhw,deSalas:2020pgw,Esteban:2020cvm}. In spite of this, the two data sets are statistically consistent.


Several approaches have been proposed to explain neutrino data~\cite{reviews}, being the two RH neutrino extension of the SM (2RH$\nu$SM)~\cite{King:1999mb,Ibarra:2003up} a framework often considered~\cite{Gautam:2015kya,Li:2017zmk,Shimizu:2017fgu,Samanta:2017kce,Nath:2018xih,Nath:2018hjx,Shimizu:2018pck,Xing:2020ald}, also in the context of leptogenesis~\cite{GonzalezFelipe:2003fi,Joaquim:2005zv,Abada:2006fw,Nardi:2006fx,Guo:2006qa,Blanchet:2011xq,Bjorkeroth:2015tsa,Bambhaniya:2016rbb,Siyeon:2016wro,Achelashvili:2017nqp,Shimizu:2017vwi,Zhao:2020bzx}. Following the Occam's razor spirit, previous studies have shown that the 2RH$\nu$SM with texture-zero Yukawa and mass matrices is not compatible with data for a NO spectrum~\cite{Frampton:2002qc,Guo:2003cc,Harigaya:2012bw,Zhang:2015tea,Rink:2016vvl,Barreiros:2018ndn,Barreiros:2018bju}. Moreover, for leptogenesis to be effective in generating the observed value of the BAU, the lightest RH neutrino mass $M_1$ should lie close to $10^{14}$~GeV~\cite{Barreiros:2018ndn}. Since, typically, the reheating temperature of the Universe obeys $T_{\rm Rh}\gtrsim M_1$~\cite{Ellis:2003sq,Giudice:2003jh}, that value of $M_1$ is in conflict with vanilla scenarios where the DM density is due to a Peccei-Quinn (PQ) axion~\cite{Peccei:1977hh} with $f_a \simeq 10^{12}$~GeV~\cite{Preskill:1982cy,Abbott:1982af,Dine:1982ah}. Indeed, if $T_{\rm Rh} > f_a$, the PQ symmetry is restored, even if broken before or during inflation. The required spontaneous breaking of that symmetry after reheating would lead to domain-wall production, which is incompatible with standard cosmology~\cite{Sikivie:1982qv} (this is, however, not a problem in a special axion model with domain-wall number $N_{\rm DW}=1$~\cite{Vilenkin:1982ks,Kawasaki:2014sqa}). So, besides being incompatible with NO neutrino masses, the minimal seesaw with Occam's razor studied so far is also in tension with the axion DM hypothesis. Notice, however, that to keep the number of parameters as minimal as possible, charged leptons were assumed to be in their mass basis, in which case lepton mixing originates solely from the neutrino sector.

Guided by the same Occam's razor principle, we show in this work that a NO neutrino mass spectrum and $\delta$ lying in the experimentally allowed region can be simultaneously accounted for with the same number of parameters as in the aforementioned studies, but with nontrivial charged-lepton mixing parametrized by a single angle $\theta_\ell$. The atmospheric neutrino mixing angle $\theta_{23}$ is predicted to be in the second octant and, interestingly, $\theta_\ell$ is very close to the quark Cabibbo angle $\theta_{\rm C} \simeq 0.23$. Moreover, taking the present best-fit value for the low-energy CP-violating phase $\delta$, successful generation of the BAU through (flavored) leptogenesis requires the lightest RH neutrino mass to be $\sim 5.5\times 10^{10}$~GeV, well below the {\em problematic}  $10^{14}$~GeV value referred above.

Let us start by considering the 2RH$\nu$SM in which the minimal type I seesaw can be realized. The Lagrangian for the lepton Yukawa and mass terms reads
\begin{gather}
\hspace{-0.2cm}\mathcal{L}=-\overline{\ell_{L}}\Ynu^* \tilde{\Phi}\nu_{R} -\frac{1}{2}\overline{\nu_{R}}\,\MR\nu_{R}^c -\overline{\ell_{L}}\Yl \Phi\,e_{R}  + \text{H.c.}\,,\label{lagrangian}
\end{gather}
where $\Ynu$ ($\Yl$) is the $3\times 2$ ($3\times 3$) Dirac neutrino (charged-lepton) Yukawa coupling matrix, and $\MR$ is the symmetric $2\times 2$ RH neutrino mass matrix. As usual, $\Phi=(\phi^+\,\, \phi^0)^T$ stands for the SM Higgs doublet with $\tilde{\Phi}=i\sigma_2 \Phi^\ast$ ($\sigma_2$ is the complex Pauli matrix in the usual notation). The left-handed (LH) lepton doublets, RH charged-lepton and RH neutrino singlets are denoted by $\ell_L=(\nu\,\,e)_L$, $e_R$ and $\nu_R$, respectively. The charged-lepton and seesaw neutrino mass matrices generated upon electroweak symmetry breaking are
\begin{align}
\Mell=v\Yl\;,\;\;\Mnu=-v^2\Ynu\MR^{-1}\Ynu^T
\,,
\label{seesaw}
\end{align}
being $v \simeq 174$~GeV the vaccuum expectation value of $\phi^0$. The relevant LH field rotations to the mass basis will be denoted by the unitary matrices $\Unu$ and $\Ul$ such that
\begin{align}
\Unu^T\Mnu\Unu&=\text{diag}(m_1,m_2,m_3)\equiv \mathbf{d}_m
\,,\label{Mnudiag}\\ 
\Ul^\dagger\Mell\Mell^\dagger\Ul&=\text{diag}(m_e^2,m_\mu^2,m_\tau^2)\,,
\end{align}
where, $m_{1,2,3}$ and $m_{e,\mu,\tau}$ are the light-neutrino and charged-lepton masses, respectively. The lepton mixing matrix probed by neutrino oscillation experiments is $\U=\Ul^\dagger\Unu^{}$. We recall that, since $m_1=0$ in the 2RH$\nu$SM, $m_2^2=\dmsol$ and $m_3^2=\dmatm$, being the neutrino mass-squared differences $\Delta m^2_{21,31}$ also measured experimentally.

In the charged-lepton mass basis ($\Ul=\openone$), the most restrictive texture-zero matrices $\Ynu$ and $\MR$ (labelled as T$_{ab}$ and $\text{R}_{cd}$, respectively) compatible with data are~\cite{Barreiros:2018ndn}:
\begin{align}
\text{T}_{ab}: &\,(\Ynu)_{a1}=(\Ynu)_{b2}=0\,,\; a\neq b=1,2,3\,,\label{Ynutextures}\\
\text{R}_{cd}: &\,(\MR)_{cd}=0\,,\;c\leq d=1,2\,.\label{MRtextures}
\end{align}
The effective neutrino mass matrix $\Mnu$ generated by the seesaw formula given in eq.~\eqref{seesaw} with any of the ($\Ynu$,$\MR$)=(T$_{ab}$,R$_{cd}$) pairs always exhibits one of the following structures
\begin{align}
\hspace{-0.20cm}\begin{matrix}
\text{A}:\begin{pmatrix}
0&\times&\times\\
\cdot&\times&\times\\
\cdot&\cdot&\times
\end{pmatrix},&
\hspace{-0.1cm}\text{B}:\begin{pmatrix}
\times&0&\times\\
\cdot&\times&\times\\
\cdot&\cdot&\times
\end{pmatrix},&
\hspace{-0.1cm}\text{C}:\begin{pmatrix}
\times&\times&0\\
\cdot&\times&\times\\
\cdot&\cdot&\times
\end{pmatrix},\\\\[-0.18cm]
\text{D}:\begin{pmatrix}
\times&\times&\times\\
\cdot&0&\times\\
\cdot&\cdot&\times
\end{pmatrix},&
\hspace{-0.1cm}\text{E}:\begin{pmatrix}
\times&\times&\times\\
\cdot&\times&0\\
\cdot&\cdot&\times
\end{pmatrix},&
\hspace{-0.1cm}\text{F}:\begin{pmatrix}
\times&\times&\times\\
\cdot&\times&\times\\
\cdot&\cdot&0
\end{pmatrix}.
\end{matrix}\label{Mnutextures}
\end{align}
From eq.~\eqref{Mnudiag}, and taking into account that $\Unu=\Ul \U$, the low-energy constraint imposed by the condition $(\Mnu)_{ij}=0$ for each texture ${\rm A}-{\rm F}$ reads
\begin{align}
\dfrac{m_2}{m_3}=-\dfrac{(\mathbf{U}_\ell^{*}\U^*)_{i3}(\mathbf{U}_\ell^{*}\U^*)_{j3}}{(\mathbf{U}_\ell^{*}\U^*)_{i2}(\mathbf{U}_\ell^{*}\U^*)_{j2}}\,.
\label{mastereq}
\end{align}
As usual, we express $\U=\V\,\text{diag}(1,e^{i\alpha/2},1)$, where $\V$ is defined by the three mixing angles $\theta_{ij}$ $(i<j=1,2,3)$ and the Dirac-type phase $\delta$ (here, we consider $\V$ parametrized as in ref.~\cite{Barreiros:2018bju}). Since $m_1=0$, there is a single Majorana phase $\alpha$.  With this convention, eq.~\eqref{mastereq} leads to the low-energy relations
\begin{align}
r_\nu&=\sqrt{\frac{\Delta m_{21}^2}{\Delta m_{31}^2}}=\left|\dfrac{(\mathbf{U}_\ell^{*}\V^*)_{i3}(\mathbf{U}_\ell^{*}\V^*)_{j3}}{(\mathbf{U}_\ell^{*}\V^*)_{i2}(\mathbf{U}_\ell^{*}\V^*)_{j2}}\right|\,,\label{rnudef}\\
\alpha&=-\arg\left[-\dfrac{(\mathbf{U}_\ell^{*}\V^*)_{i3}(\mathbf{U}_\ell^{*}\V^*)_{j3}}{(\mathbf{U}_\ell^{*}\V^*)_{i2}(\mathbf{U}_\ell^{*}\V^*)_{j2}}\right]\label{alphadef}\,.
\end{align}
It has been shown that in the 2RH$\nu$SM {\em none of the textures given in eq.~\eqref{Mnutextures} is compatible with present neutrino oscillation data for a NO spectrum when all lepton mixing stems from the neutrino sector, i.e. when} $\Ul=\openone$~\cite{Barreiros:2018ndn}.

In the Occam's razor spirit, we will consider simple cases of two-flavour mixing in the charged-lepton sector, with the same number of parameters as when $\Mell$ is diagonal, i.e. three! Namely, the textures for $\Mell$ to be studied are 
\begin{gather}
\text{L}_1^k: \begin{pmatrix}
m_k&0&0\\
0&0&\epsilon\\
0&\epsilon& m
\end{pmatrix}\!,\,
\text{L}_2^k: \begin{pmatrix}
0&0&\epsilon\\
0&m_k&0\\
\epsilon&0& m
\end{pmatrix}\!,\,
\text{L}_3^k: \begin{pmatrix}
0&\epsilon&0\\
\epsilon &m &0\\
0&0&m_k
\end{pmatrix}\!\!,\label{Mltextures}
\end{gather}
where the charged lepton with flavour $k=e,\mu,\tau$ is decoupled.  In each case, $\epsilon$ and $m$ are determined by the charged-lepton masses as $\epsilon=\pm\sqrt{m_i m_j}$ and $m=m_j-m_i$ with $i\neq j\neq k=e,\mu,\tau$ and $m_j>m_i$. The corresponding $2\times 2$ rotation in the $(i,j)$ plane which diagonalizes L$_{1,2,3}^k$ is defined by the angle $\theta_\ell^k$ 
\begin{gather}
\theta_\ell^k=\pm\dfrac{1}{2}\arctan\left(\dfrac{2\sqrt{m_i m_j}}{m_j-m_i}\right) \simeq  \pm\sqrt{m_i/m_j}\,,
\label{thetalk}
\end{gather}
which, using the experimental values of $m_{e,\mu,\tau}$, implies: 
\begin{align}
|\theta_\ell^e| \simeq 0.24\,\,,\,\,
|\theta_\ell^\mu| \simeq  0.017\,\,,\,\,
|\theta_\ell^\tau| \simeq  0.07\,.
\end{align}
Notice that $\theta_\ell^e\simeq \sqrt{m_\mu/m_\tau}\simeq 0.24$ is close to the quark Cabibbo angle $\theta_{\rm C}\simeq 0.23$. As it will be shortly seen, the case ${\rm L}_1^e$ turns out to be preferred by data. From now on we will only focus on textures L$_1^e$, L$_2^\mu$ and L$_3^\tau$ for $\Mell$ since the remaining possibilities are related to those by row and column permutations. Any $i$-row permutation must be also performed in $\Ynu$, implying a simultaneous $i$-row and -column permutation in $\Mnu$. Thus, some pairs $(\Mnu$,$\mathbf{M}_\ell)=({\rm A-F},{\rm L}_{1,2,3}^k)$, with A--F and ${\rm L}_i^k$ given in \eqref{Mnutextures} and \eqref{Mltextures}, respectively, are actually equivalent with respect to their low-energy predictions. 

\begin{figure}[t!]
	\centering
	\begin{tabular}{l}
	\includegraphics[width=0.48\textwidth]{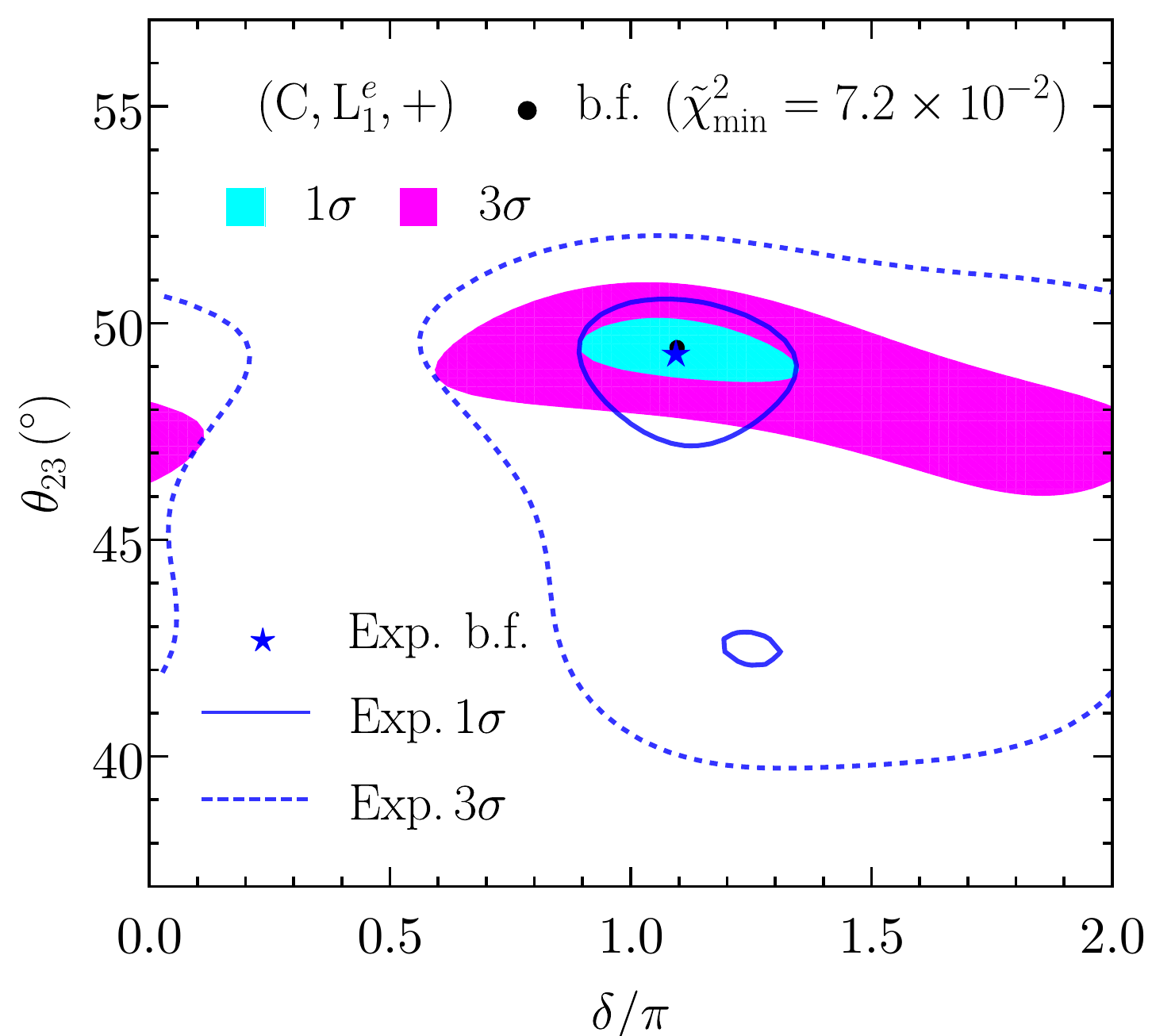} \\
	\includegraphics[width=0.48\textwidth]{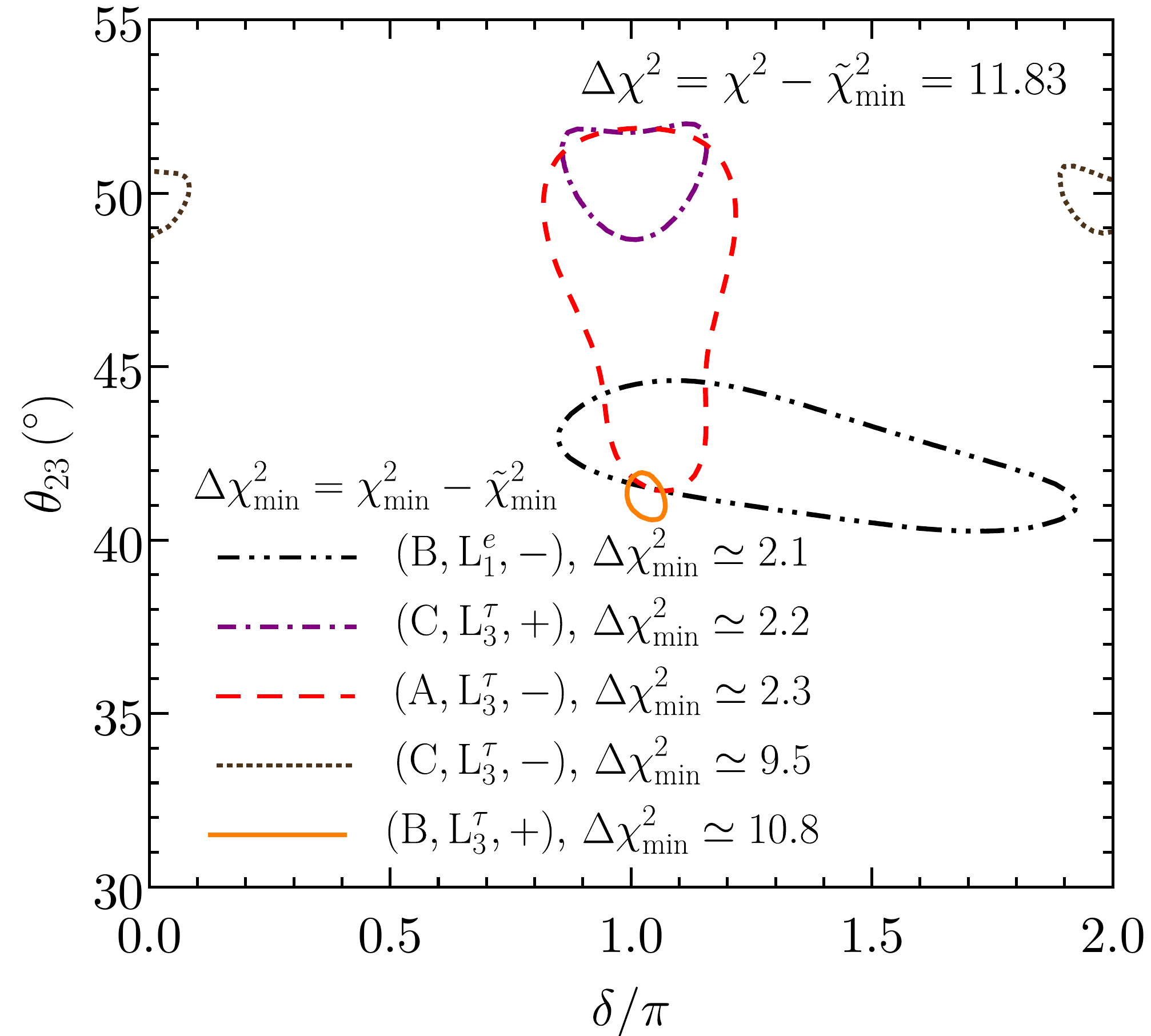}
	\end{tabular}
\caption{ Top: $1\sigma$ and $3\sigma$ allowed regions in the $(\delta,\theta_{23})$ plane are shown in cyan and magenta, respectively, for the best-fit combination ($\Mnu$,$\Mell$,$\theta_\ell^e$)=(C,${\rm L}_1^e$,+). The point marked with a black dot corresponds to the absolute minimum $\tilde{\chi}^2_{\rm min}=7.2\times10^{-2}$. For comparison, the lines delimiting the regions allowed by experimental data only are also shown (with the best-fit point marked with a blue star). Bottom: Delimiting lines of the $3\sigma$ regions ($\Delta \chi^2=11.83$) and corresponding values of $\Delta \chi^2_{\rm min}=\chi^2_{\rm min}-\tilde{\chi}^2_{\rm min}$ for the remaining texture combinations.}
	\label{t23pred}
\end{figure} 

In order to test the compatibility of the new Occam's razor setups with neutrino data we proceed as follows. For each $(\Mnu$,$\mathbf{M}_\ell,\theta_\ell^k)=({\rm A-F},{\rm L}_{1,2,3}^k,\pm)$ case, where $\pm$ indicates the sign of $\theta_\ell^k$ in eq.~\eqref{thetalk}, we obtain the corresponding low-energy constraint defined by eqs.~\eqref{rnudef} and \eqref{thetalk}. Since the atmospheric mixing angle $\theta_{23}$ is currently the less precisely measured oscillation parameter (besides $\delta$), we choose to express $\theta_{23}$ as function of $\theta_{12,13}$, $\Delta m^2_{21,31}$, $\delta$ and $\theta_\ell^k$. We then minimize the total chi-squared function $\chi^2_{\rm tot}$ using the one-dimensional $\chi^2$ distributions given in  refs.~\cite{Esteban:2020cvm,globalfit} for all oscillation parameters, except for $\theta_{23}$ and $\delta$. Their contribution to $\chi^2_{\rm tot}$ is instead determined using an interpolating function of the two-dimensional $\chi^2(\theta_{23},\delta)$ data sample publicly available at~\cite{globalfit}. Obviously, the value of $\theta_{23}$ is the one implied by Occam's razor and extracted from \eqref{rnudef}. After identifying the best-fit scenario and the corresponding $\tilde{\chi}^2_{\rm min}$, we find the $1\sigma$ and $3\sigma$ allowed regions in the $(\delta,\theta_{23})$ plane for all cases compatible with data which correspond to $\Delta \chi^2=\chi^2-\tilde{\chi}^2_{\rm min}\leq 2.30$ and 11.83, respectively. 
\begin{figure}[t!]
	\centering
	\begin{tabular}{l}
	\includegraphics[width=0.48\textwidth]{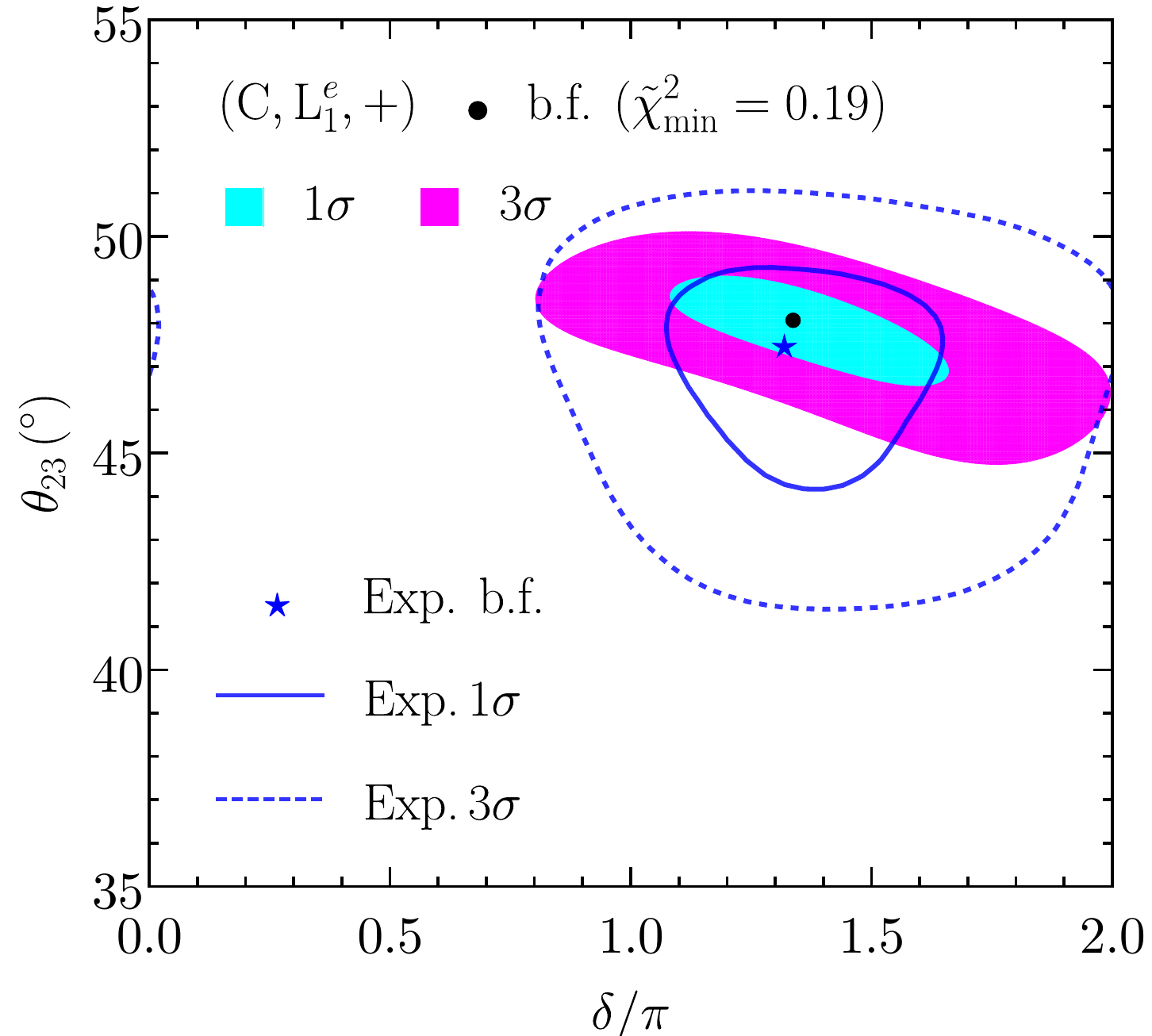} \\
	\includegraphics[width=0.48\textwidth]{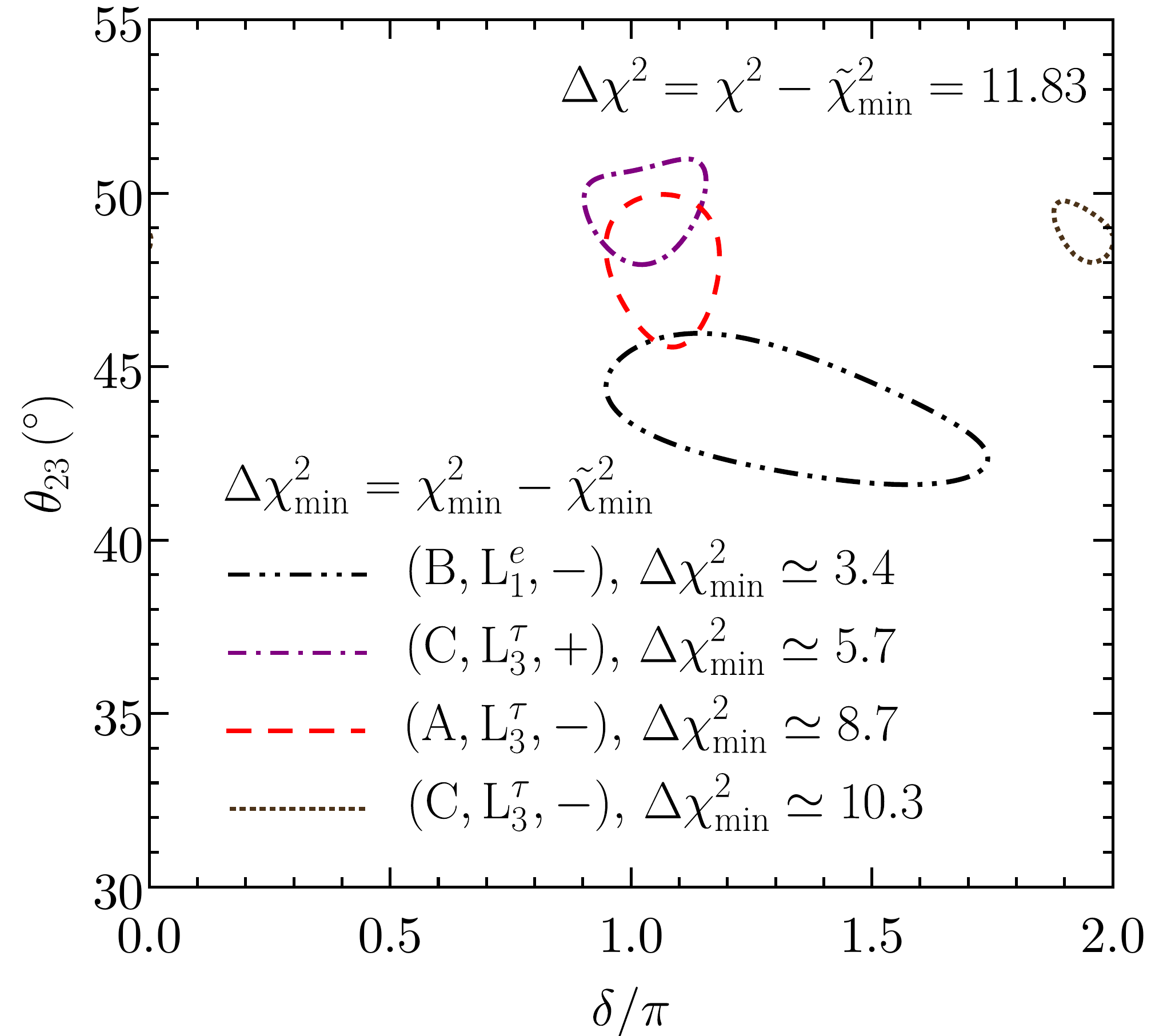}
	\end{tabular}
	\caption{For comparison, we show in this figure the same as in fig.~\ref{t23pred}, but considering the pre-Neutrino 2020 global-fit of Refs.~\cite{deSalas:2017kay,globalfit2}, i.e. without the new NO$\nu$A and T2K data~\cite{Acero:2019ksn,Abe:2019vii}.}
	\label{t23predold}
\end{figure} 

Our results show that the best scenario is the one provided by the combination (C,L$_1^e$,+) with $\tilde{\chi}^2_{\rm min}=7.2\times 10^{-2}$. The $1\sigma$ and $3\sigma$ regions are presented in the top panel of fig.~\ref{t23pred}, where the experimental regions are also shown for comparison. Clearly, $\theta_{23}$ is predicted to be in the second octant. Other combinations like (B,L$_1^e$,$-$), (C,L$_3^\tau$,$+$) and (A,L$_3^\tau$,$-$) are marginally compatible with data at $1\sigma$ while (C,L$_3^\tau$,$-$) and (B,L$_3^\tau$,$+$) are only compatible at $3\sigma$, as shown in the bottom panel of the same figure. Notice that, except for the case (A,L$_3^\tau$,$-$), all those solutions select one of the $\theta_{23}$ octants. Thus, future clarification of the $\theta_{23}$ octant problem will be essential to scrutinize the Occam's razor hypothesis put forward in this work, rending it testable. It is also worth emphasizing that for the preferred case $\Mell$ is of L$_1^e$ type and, thus, the mixing coming from the charged lepton sector is very close to quark Cabibbo mixing. To evaluate the impact of the latest T2K~~\cite{Acero:2019ksn} and NO$\nu$A~~\cite{Abe:2019vii} data, we present in fig.~\ref{t23predold} the same as in fig.~\ref{t23pred} but considering the pre-Neutrino2020 global fit results of Refs.~\cite{deSalas:2017kay,globalfit2}, which do not include that data. By comparing the two figures we conclude that after including the new T2K and NO$\nu$A data (C,L$_1^e$,+) remains the preferred scenario with a better fit quality, being (B,L$_1^e$,$-$), (C,L$_3^\tau$,$-$) and (A,L$_3^\tau$,$-$) now marginally compatible with data at $1\sigma$.

\begin{figure}[t!]
	\centering
	\begin{tabular}{l}
	\includegraphics[width=0.48\textwidth]{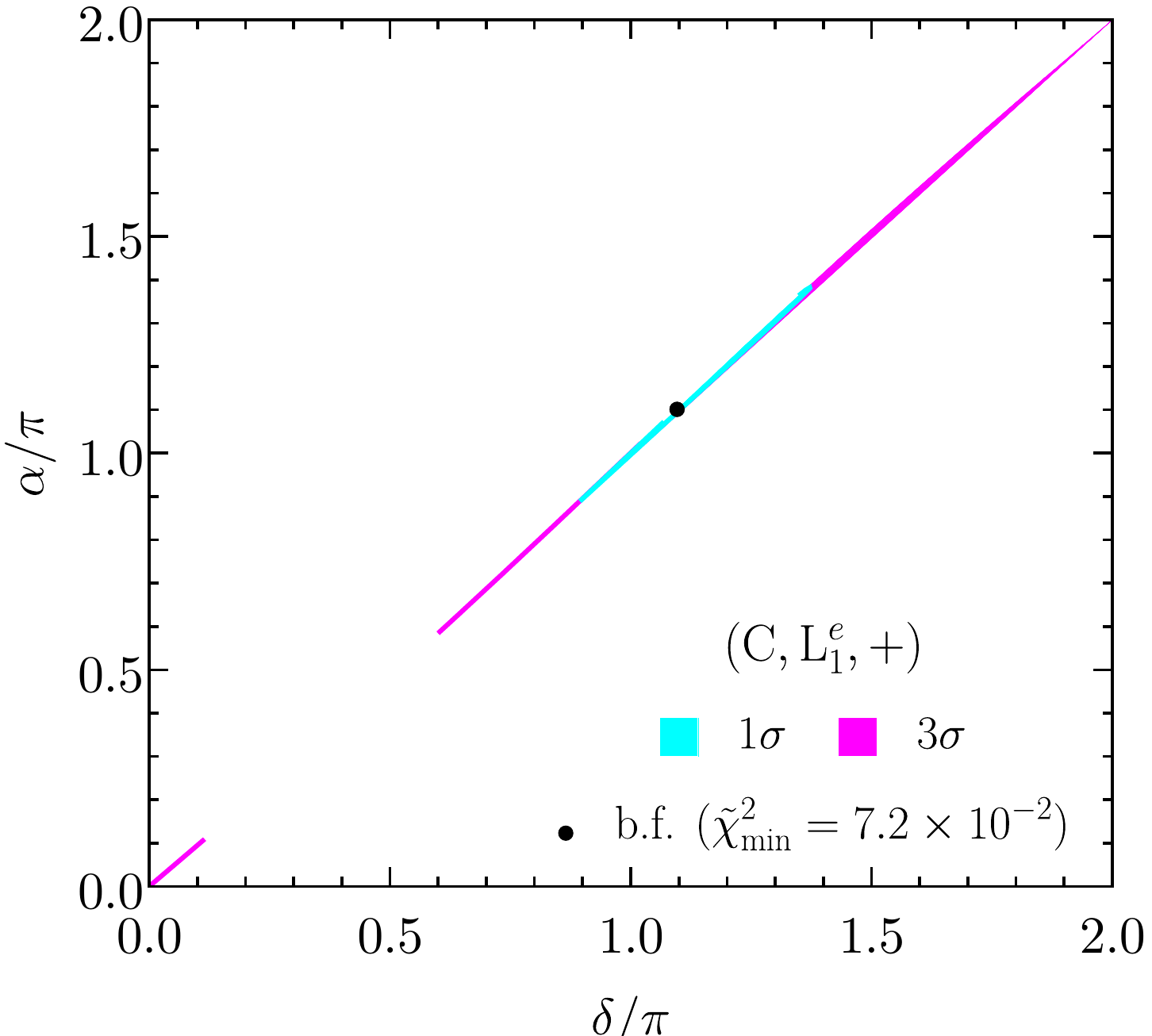} \\
	\includegraphics[width=0.48\textwidth]{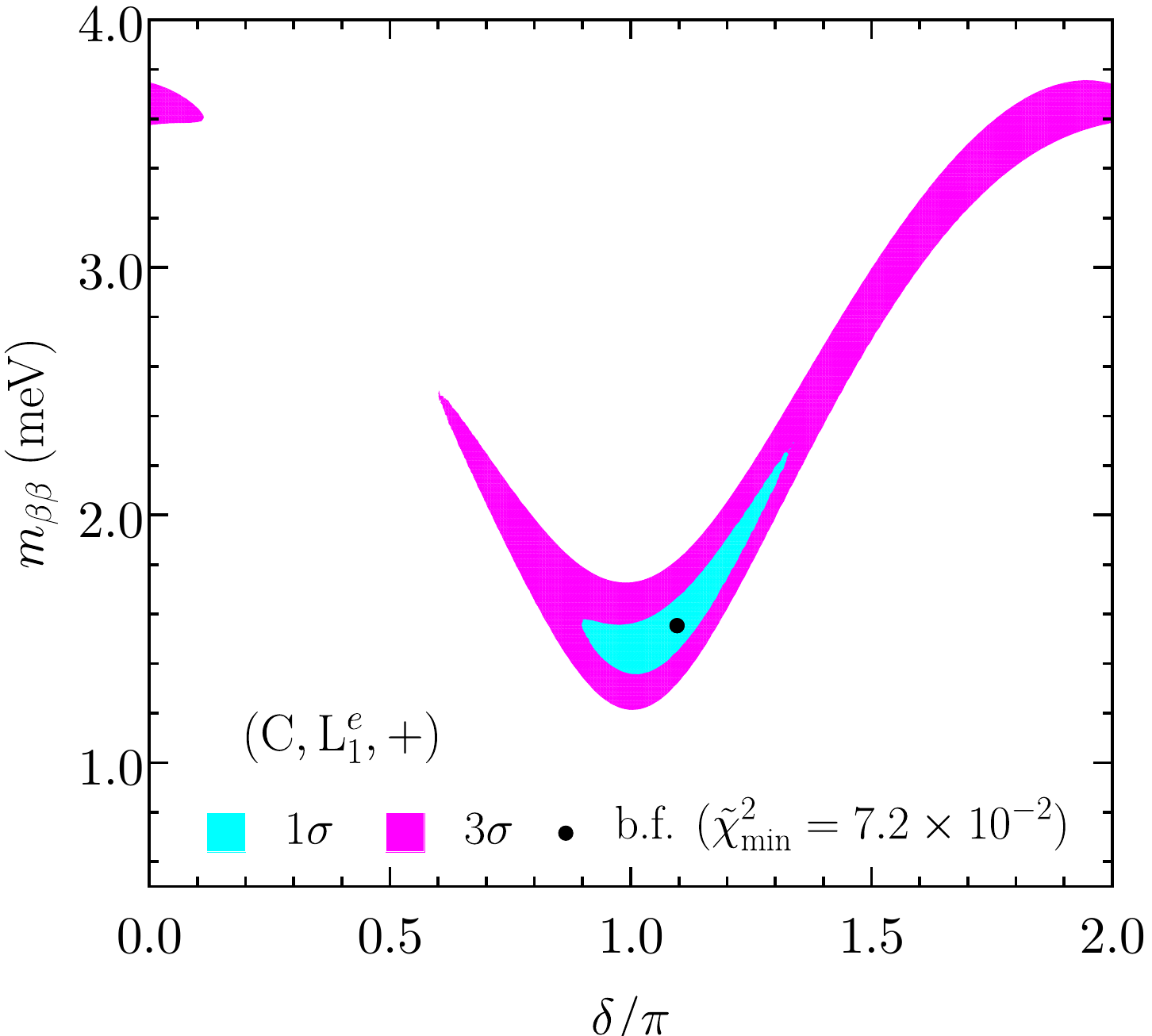}
	\end{tabular}
	\caption{Majorana CP-violating phase $\alpha$ (top panel) and effective neutrino mass parameter $m_{\beta\beta}$ (bottom panel) as function of the Dirac CP phase $\delta$.}
	\label{mbbfig}
\end{figure}

For (C,L$_1^e$,+) and (B,L$_1^e$,$-$), the dependence of $\theta_{23}$ on the remaining oscillation parameters and on the charged-lepton mixing angle $\theta_\ell^e$ can be expressed by 
\begin{align}
\!\!\tan(\theta_{23}+\theta_\ell^e)&\simeq\left[\dfrac{2s_{13}\mp 2r_\nu s_{13}s_{12}^2c_\delta   }{r_\nu\sin(2\theta_{12})}\right]^{\pm 1}
\!\!\!,\;
\alpha\simeq \pm \,\delta\,, 
\label{LEcondB3e}
\end{align}
which hold to a very good approximation (we use the notation $c_{ij}\equiv\cos\theta_{ij} $ and $s_{ij}\equiv\sin\theta_{ij}$). The upper and lower signs apply to the (C,L$_1^e$,+) and (B,L$_1^e$,$-$) combinations, respectively. Notice that the Majorana phase $\alpha$ is directly related to the Dirac CP phase $\delta$. In the top panel of fig.~\ref{mbbfig}, the allowed regions in the $(\delta,\alpha)$ plane are shown for (C,L$_1^e$,+), confirming the approximate result $\alpha \simeq \delta$ in \eqref{alphadef}. Using relations \eqref{rnudef} and \eqref{alphadef}, the dependence of $m_{\beta\beta}$ (the effective neutrino mass parameter relevant for neutrinoless beta decay) and $\delta$ can be established, as shown in the bottom panel of fig.~\ref{mbbfig} for (C,L$_1^e$,+). The results show that $m_{\beta\beta} \in [1.2,3.8]$~meV at $3\sigma$, being 1.55~meV at the best-fit point.

Besides setting the minimal seesaw template for neutrino mass generation, the 2RH$\nu$SM also provides the most economical framework for the generation of the BAU through the leptogenesis mechanism~\cite{Fukugita:1986hr}. In this context, the baryon-to-photon ratio $\eta_B$ is given by~\cite{Buchmuller:2004nz}
\begin{gather}
\eta_B=a_\text{sph}\,N_{B-L}/N_\gamma^\text{rec} \simeq 9.6\times 10^{-3}\, N_{B-L}\;  ,
\label{finaletab}
\end{gather}
being $a_\text{sph}=28/79$ the sphaleron conversion factor, $N_{B-L}$ the final $B-L$ asymmetry and $N_\gamma^\text{rec}\simeq 37.0$ the number of photons at the recombination temperature computed in the same comoving volume. The present $\eta_B$ value determined from Planck Collaboration data is~\cite{Ade:2015xua} 
 \begin{align}
 \eta_B^0=(6.11\pm0.04)\times 10^{-10}.
 \label{etabexp}
\end{align}

The generation of $\eta_B$ through leptogenesis requires non-vanishing CP asymmetries in the RH neutrino decays $\nu_{Ri}\rightarrow \ell_\alpha \Phi$, given by~\cite{Covi:1996wh,Abada:2006ea}
\begin{align}
\epsilon_i^\alpha =&\frac{1}{8\pi}\frac{1}{\mathbf{H}_{ii}^\nu}\{\text{Im}[(\Ynut^*)_{\alpha i}\mathbf{H}_ {ij}^\nu (\Ynut)_{\alpha j}][f(x_j)+g(x_j)]\nonumber\\
&+\text{Im}[(\Ynut^*)_{\alpha i}\mathbf{H}_ {ji}^\nu(\Ynut)_{\alpha j}]g'(x_j)\},
\label{flavouredcp}	
\end{align}
where $j\neq i=1,2$, $x_j=M_j^2/M_i^2$, being $M_i$ the heavy neutrino masses. Here, $\Ynut=\Ul^T \Ynu$ denotes $\Ynu$ in the charged-lepton diagonal basis,  $\mathbf{H}^\nu=\mathbf{Y}^{\nu\dagger} \mathbf{Y}^\nu$, and $f(x)$, $g(x)$ and $g'(x)$ are loop functions (see e.g. ref.~\cite{Branco:2011zb}). To compute $N_{B-L}$ we consider two regimes differing from their $M_{1,2}$ validity ranges~\cite{Barbieri:1999ma,Abada:2006ea}: the flavored regime with $10^9\lesssim M_{1,2}\lesssim10^{12}$~GeV (only $\tau$ Yukawa interactions are in equilibrium), and the unflavored regime with $M_{1,2}\gtrsim10^{12}$~GeV (all Yukawa interactions are out of equilibrium, being indistinguishable). In the former case, the relevant CP asymmetries are $\epsilon_i^\tau$ and $\epsilon_i^\gamma=\epsilon_i^e+\epsilon_i^\mu$, while in the latter only the asymmetry summed over all flavors, i.e. $\epsilon_i=\sum_\alpha\epsilon_i^\alpha$, needs to be considered. The efficiency factors in the production of $N_{B-L}$ ($\kappa_i^\alpha$ and $\kappa_i$ for the flavored and unflavored regimes, respectively) can be computed using the results of refs.~\cite{Buchmuller:2004nz} and \cite{Antusch:2011nz}. 

In the remaining, we address the question of whether the observed BAU can be generated via leptogenesis in the preferred Occam's razor scenario with ($\Mnu$,$\Mell$,$\theta_\ell^e$)=(C,L$_1^e$,+). Within the 2RH$\nu$SM with texture-zero $\Ynu$ and $\MR$, a neutrino mass matrix $\Mnu$ of the type C shown in \eqref{Mnutextures} can only be generated by combining $\Ynu$ and $\MR$ matrices of the type T$_{13,31}$ and R$_{12}$ defined in \eqref{Ynutextures} and \eqref{MRtextures}, respectively. To compute the CP asymmetries $\epsilon_i^\alpha$, we use the Casas-Ibarra parametrization~\cite{Casas:2001sr}
\begin{align}
\mathbf{Y}^\nu=v^{-1}\Ul^*\U^*\,\mathbf{d}_m^{1/2}\,\mathbf{R}\,\mathbf{d}_M^{1/2}\,,
\label{CasasandIbarra}
\end{align}
where $\mathbf{d}_m$ has been defined in eq.~\eqref{seesaw} and $\mathbf{d}_M=\text{diag}(M_1,M_2)$. Since $m_1=0$, the complex orthogonal $3\times 2$ matrix $\mathbf{R}$ can be parametrized by a single complex angle $z$ and a discrete-valued parameter $\xi=\pm 1$~\cite{Ibarra:2003up}. The conditions for $\Ynu$ given in \eqref{Ynutextures} determine $z$ in terms of low-energy parameters. Namely, for a NO spectrum,
\begin{align}
(\Ynu)_{i1}=0:\,&
\tan z=-\xi \sqrt{\dfrac{m_2}{m_3}}\dfrac{(\Unu^*)_{i2}}{(\Unu^*)_{i3}}\,,\,
\label{y11=0}\\
(\Ynu)_{i2}=0:\,&
\tan z=\xi \sqrt{\dfrac{m_3}{m_2}}\dfrac{(\Unu^*)_{i3}}{(\Unu^*)_{i2}}\,,
\label{y12=0}
\end{align}
which, together with the low-energy relations given in eq.~\eqref{LEcondB3e}, determine the CP asymmetries and corresponding efficiency factors $\kappa_i$. The results when $\Ynu$ is of the type T$_{13}$ may be obtained from those with T$_{31}$ through $M_1\leftrightarrow M_2$, i.e. $r_N\rightarrow 1/r_N$ where $r_N=M_2/M_1$. It can also be shown that $\epsilon_i^e=0$, and that the dependence of $\kappa_i$ on $M_{1,2}$ can be safely neglected. 

In the following, all presented numerical results were obtained considering the values of the neutrino parameters which minimize $\chi^2_{\rm tot}$ for ($\Mnu$,$\Mell$,$\theta_\ell^e$)=(C,L$_1^e$,+), i.e. those corresponding to the black dot in the left panel of fig.~\ref{t23pred}. We obtain $\eta_B<0$ for texture T$_{13}$ and $r_N>1$, in both flavored and unflavored regimes. Therefore, these cases are obviously excluded and, from now on, we will restrict ourselves to T$_{31}$ with $r_N>1$. In the unflavored regime ($M_1\gtrsim10^{12}$~GeV), the dominant contribution to $\eta_B$ comes from $\nu_{R1}$ decays implying $\eta_B\simeq-9.6\times 10^{-3}\kappa_1\epsilon_1$ with $\kappa_1 \simeq 3.7\times 10^{-3}$. Taking $\alpha\simeq \delta$, as obtained in \eqref{alphadef}, 
\begin{align}
\epsilon_1\simeq\dfrac{3M_1 \sqrt{\dmatm} s_{13}^2s_\delta}{16\pi v^2(s_{13}^2+r_\nu^2 s_{12}^2 c_{13}^2)}\,,
\end{align} 
leading to $\eta_B \simeq 6.9 \times 10^{-22} (M_1/{\rm GeV)}$ which, for $M_1 \gtrsim 10^{12}$~GeV, implies $\eta_B \gtrsim 6.9\times 10^{-10}\simeq 1.1\,\eta_B^0$. Including the contribution of the heaviest RH neutrino one gets $\eta_B \gtrsim 1.2\times 10^{-9}\simeq 1.9\,\eta_B^0$. Thus, $\eta_B$ is too large and unflavored leptogenesis is disfavored.
\begin{figure}[t!]
	\begin{tabular}{c}
	\includegraphics[width=0.48\textwidth]{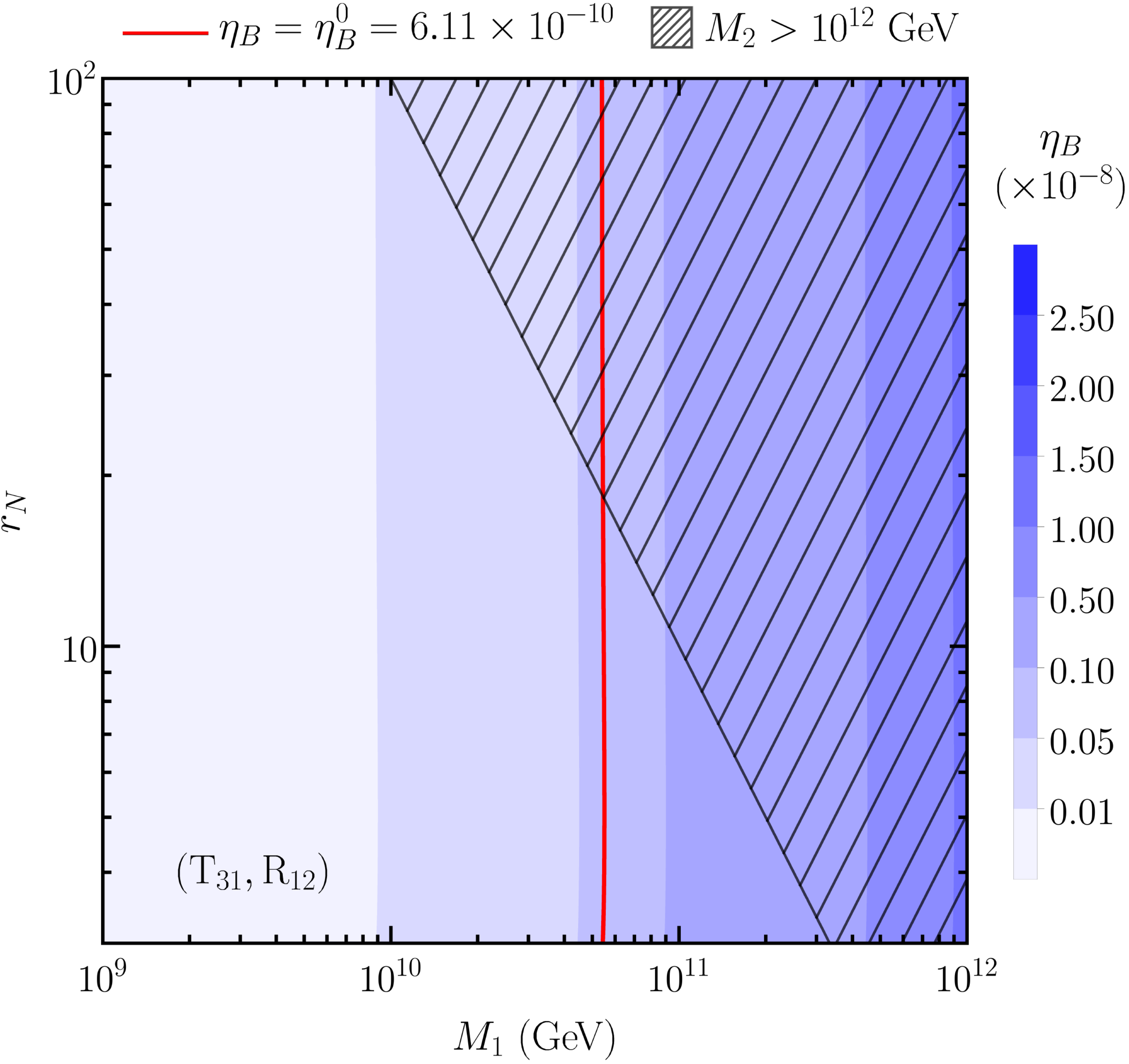}
	\end{tabular}
\caption{$\eta_B$ contours in the plane ($M_1$,$r_N$) obtained with a full numerical analysis based on the results of ref.~\cite{Antusch:2011nz} for flavored leptogenesis applied to the case ($\Mnu$,$\Mell$,$\theta_\ell^e$)=(C,L$_1^e$,+) built from ($\Ynu$,$\MR$)=(T$_{31}$,R$_{12}$). The low-energy constraints corresponding to texture C for $\Mnu$ and the best-fit values for the oscillation parameters have been used. In the hatched region $M_2$ is out of the mass interval for the flavored leptogenesis regime. The red solid contour corresponds to the experimental value \eqref{etabexp}.}
\label{etabflvpred}
\end{figure} 

Turning now to the flavored regime, and assuming $M_2\gtrsim 3 M_1$ as in ref.~\cite{Antusch:2011nz}, we checked that the dominant contributions to $\eta_B$ come from $\epsilon_1^{\tau,\gamma}$, being $N_{B-L}\simeq-\kappa_1^\tau\epsilon_1^\tau-\kappa_1^\gamma\epsilon_1^{\gamma}$ with
 \begin{align}
\epsilon_1^\tau&\simeq\dfrac{3M_1\sqrt{\dmatm}s_\delta s_{13}(s_{13}c_{23}^2-r_\nu s_{12}c_{12}s_{23}c_{23})}{16\pi v^2(r_\nu^2 s_{12}^2+s_{13}^2)}\nonumber\,,\\
 \epsilon_1^\gamma&= \epsilon_1^\mu \simeq \dfrac{3M_1\sqrt{\dmatm}s_\delta s_{13} s_{23}c_{23}}{16\pi v^2r_\nu \tan\theta_{12}}\,,
 \end{align}
where $\alpha\simeq \delta$ was again used. Given that $\kappa_1^\tau\simeq 1.7\times 10^{-1}$ and $\kappa_1^\gamma\simeq 5.6\times 10^{-3}$, we obtain $\eta_B \simeq 6.0\times 10^{-21}(M_1/{\rm GeV})$.
Thus, the experimental value of $\eta_B$ requires $M_1\sim 1\times 10^{11}$~GeV. In fig.~\ref{etabflvpred}, we show the results for $\eta_B$ obtained with a full numerical analysis, considering the contributions of all flavours in the decays of both heavy neutrinos. The red contour for $\eta_B=\eta_B^0$ at $M_1 = 5.5\times 10^{10}$~GeV is in agreement with the estimate above.\footnote{With the pre-Neutrino2020 global-fit results we obtain $M_1 = 2.2\times 10^{10}$~GeV.} It is also clear from the figure that $\eta_B$ is not sensitive to $M_2$.
\vspace*{0.3cm}

In conclusion, the new Occam's razor scenario for texture-zero Yukawa and mass matrices with ($\Mnu$,$\Mell$,$\theta_\ell^e$)=(C,L$_1^e$,+) is compatible with a normally-ordered neutrino mass spectrum, and predicts $\theta_{23}$ belonging to the second octant. The difference with respect to previous works, where only invertedly-ordered neutrino masses are compatible with data, is that charged-lepton mixing is introduced without increasing the number of parameters in $\Mell$. Interestingly, this mixing is predicted to be very close to quark Cabibbo mixing. Another appealing feature of this setup is that the mass of the lightest RH neutrino required for leptogenesis to work is substantially lowered from $10^{14}$~GeV (the value inferred in previous works with diagonal $\Mell$) to $5.5\times 10^{10}$~GeV. Such value is consistent with vanilla scenarios of axion dark matter in which the reheating temperature of the Universe lies below $10^{12}$~GeV to avoid harmful domain-wall production. Future neutrino oscillation results will be decisive to confirm or disprove the hypothesis put forward in this work.

\medskip
{\bf Acknowledgements:} We thank Sergio Palomares Ruiz for discussions. F.R.J. and T.T.Y. are grateful to IFT/UAM (Madrid) for the kind hospitality and financial support during the initial stage of this work. F.R.J and D.M.B acknowledge Funda\c{c}{\~a}o para a Ci{\^e}ncia e a Tecnologia (FCT, Portugal) for financial support through the projects UIDB/00777/2020, UIDP/00777/2020, CERN/FIS-PAR/0004/2017, CERN/FIS-PAR/0004/2019 and PTDC/FIS-PAR/29436/2017. The work of D.M.B. is supported by the FCT grant SFRH/BD/137127/2018. T.T.Y. is supported in part by the China Grant for Talent Scientific Start-Up Project and the JSPS Grant-in-Aid for Scientific Research No. 16H02176, No. 17H02878, and No. 19H05810 and by World Premier International Research Center Initiative (WPI Initiative), MEXT, Japan.


\begin{thebibliography}{99}


\bibitem{seesaw}
P.~Minkowski,
Phys.\ Lett.\  {\bf 67B}, 421 (1977);
%
T.~Yanagida,
Conf.\ Proc. {\bf C7902131}, 95 (1979);
M.~Gell-Mann, P.~Ramond and R.~Slansky,
Conf.\ Proc. {\bf C790927}, 315 (1979);
%
S.~L.~Glashow,
NATO Sci.\ Ser.\ B {\bf 61}, 687 (1980);
J.~Schechter and J.~W.~F.~Valle,
Phys.\ Rev.\ D {\bf 22}, 2227 (1980).

\bibitem{Fukugita:1986hr} 
  M.~Fukugita and T.~Yanagida,
  Phys.\ Lett.\ B {\bf 174}, 45 (1986).
  
\bibitem{Frampton:2002qc}
  P.~H.~Frampton, S.~L.~Glashow and T.~Yanagida,
  Phys.\ Lett.\ B {\bf 548} (2002) 119.
  
  
\bibitem{Capozzi:2020qhw} 
  F.~Capozzi, E.~Di Valentino, E.~Lisi, A.~Marrone, A.~Melchiorri and A.~Palazzo,
  arXiv:2003.08511 [hep-ph].
  
\bibitem{deSalas:2020pgw}
P.~F.~de Salas, D.~V.~Forero, S.~Gariazzo, P.~Martínez-Miravé, O.~Mena, C.~A.~Ternes, M.~Tórtola and J.~W.~F.~Valle,
[arXiv:2006.11237 [hep-ph]].
  
\bibitem{Esteban:2020cvm}
I.~Esteban, M.~C.~Gonzalez-Garcia, M.~Maltoni, T.~Schwetz and A.~Zhou,
[arXiv:2007.14792 [hep-ph]].

\bibitem{Acero:2019ksn}
M.~A.~Acero \textit{et al.} [NOvA],
Phys. Rev. Lett. \textbf{123}, no.15, 151803 (2019).

\bibitem{NovaTalk}
Himmel, A., {\em New Oscillation Results from the NOvA Experiment}. Talk given at the {\it XXIX International Conference on Neutrino Physics and Astrophysics}, Chicago, USA, June 22--July 2, 2020 (online conference).

\bibitem{Abe:2019vii}
K.~Abe \textit{et al.} [T2K],
Nature \textbf{580}, no.7803, 339-344 (2020).

\bibitem{T2KTalk}
Dunne, P., {\em Latest Neutrino Oscillation Results from T2K}. Talk given at the {\it XXIX International Conference on Neutrino Physics and Astrophysics}, Chicago, USA, June 22--July 2, 2020 (online conference).
  
\bibitem{reviews} For reviews on neutrino mass and mixing models see e.g.
  H.~Ishimori, T.~Kobayashi, H.~Ohki, Y.~Shimizu, H.~Okada and M.~Tanimoto,
  Prog.\ Theor.\ Phys.\ Suppl.\  {\bf 183}, 1 (2010);
  %
  S.~Morisi and J.~W.~F.~Valle,
  Fortsch.\ Phys.\  {\bf 61}, 466 (2013);
  S.~F.~King,
  Prog.\ Part.\ Nucl.\ Phys.\  {\bf 94}, 217 (2017);
%
  Z.~z.~Xing,
  arXiv:1909.09610 [hep-ph];
F.~Feruglio and A.~Romanino,
 arXiv:1912.06028 [hep-ph].

\bibitem{King:1999mb}
S.~F.~King,
Nucl. Phys. B \textbf{576}, 85-105 (2000),
[arXiv:hep-ph/9912492 [hep-ph]].

\bibitem{Ibarra:2003up} 
  A.~Ibarra and G.~G.~Ross,
  Phys.\ Lett.\ B {\bf 591}, 285 (2004).


\bibitem{Gautam:2015kya}
R.~R.~Gautam, M.~Singh and M.~Gupta,
Phys. Rev. D \textbf{92} (2015) no.1, 013006.

\bibitem{Li:2017zmk}
C.~C.~Li and G.~J.~Ding,
Phys. Rev. D \textbf{96} (2017) no.7, 075005.

\bibitem{Shimizu:2017fgu}
Y.~Shimizu, K.~Takagi and M.~Tanimoto,
JHEP \textbf{11} (2017), 201.

\bibitem{Samanta:2017kce}
R.~Samanta, P.~Roy and A.~Ghosal,
JHEP \textbf{06} (2018), 085.

\bibitem{Nath:2018xih}
N.~Nath,
Mod. Phys. Lett. A \textbf{34} (2019) no.39, 1950329.

\bibitem{Nath:2018hjx}
N.~Nath, Z.~z.~Xing and J.~Zhang,
Eur. Phys. J. C \textbf{78} (2018) no.4, 289.

\bibitem{Shimizu:2018pck}
Y.~Shimizu, K.~Takagi and M.~Tanimoto,
[arXiv:1805.02925 [hep-ph]].



\bibitem{Xing:2020ald}
Z.~z.~Xing and Z.~h.~Zhao,
[arXiv:2008.12090 [hep-ph]].

\bibitem{GonzalezFelipe:2003fi}
R.~Gonzalez Felipe, F.~R.~Joaquim and B.~M.~Nobre,
Phys. Rev. D \textbf{70} (2004), 085009.

\bibitem{Joaquim:2005zv}
F.~R.~Joaquim,
Nucl. Phys. B Proc. Suppl. \textbf{145} (2005), 276-279.

\bibitem{Abada:2006fw}
A.~Abada, S.~Davidson, F.~X.~Josse-Michaux, M.~Losada and A.~Riotto,
JCAP \textbf{04} (2006), 004.

\bibitem{Nardi:2006fx}
E.~Nardi, Y.~Nir, E.~Roulet and J.~Racker,
JHEP \textbf{01} (2006), 164.

\bibitem{Guo:2006qa}
W.~l.~Guo, Z.~z.~Xing and S.~Zhou,
Int. J. Mod. Phys. E \textbf{16} (2007), 1-50.

\bibitem{Blanchet:2011xq}
S.~Blanchet, P.~Di Bari, D.~A.~Jones and L.~Marzola,
JCAP \textbf{01} (2013), 041.

\bibitem{Bjorkeroth:2015tsa}
F.~Björkeroth, F.~J.~de Anda, I.~de Medeiros Varzielas and S.~F.~King,
JHEP \textbf{10} (2015), 104.

\bibitem{Bambhaniya:2016rbb}
G.~Bambhaniya, P.~S.~Bhupal Dev, S.~Goswami, S.~Khan and W.~Rodejohann,
Phys. Rev. D \textbf{95} (2017) no.9, 095016.

\bibitem{Siyeon:2016wro}
K.~Siyeon,
J. Korean Phys. Soc. \textbf{69} (2016) no.11, 1638-1643.

\bibitem{Achelashvili:2017nqp}
A.~Achelashvili and Z.~Tavartkiladze,
Nucl. Phys. B \textbf{929} (2018), 21-57.

\bibitem{Shimizu:2017vwi}
Y.~Shimizu, K.~Takagi and M.~Tanimoto,
Phys. Lett. B \textbf{778} (2018), 6-16.

\bibitem{Zhao:2020bzx}
Z.~h.~Zhao,
[arXiv:2003.00654 [hep-ph]].




  
\bibitem{Guo:2003cc} 
  W.~l.~Guo and Z.~z.~Xing,
  Phys.\ Lett.\ B {\bf 583}, 163 (2004).
  
\bibitem{Harigaya:2012bw}
  K.~Harigaya, M.~Ibe and T.~T.~Yanagida,
  Phys.\ Rev.\ D {\bf 86} (2012) 013002.

\bibitem{Zhang:2015tea}
  J.~Zhang and S.~Zhou,
  JHEP {\bf 1509} (2015) 065.
  
\bibitem{Rink:2016vvl}
  T.~Rink and K.~Schmitz,
  JHEP {\bf 1703} (2017) 158.
  
\bibitem{Barreiros:2018ndn}  
D.~M.~Barreiros, R.~G.~Felipe and F.~R.~Joaquim,
Phys.\ Rev.\ D {\bf 97} (2018) no.11,  115016.

\bibitem{Barreiros:2018bju} 
  D.~M.~Barreiros, R.~G.~Felipe and F.~R.~Joaquim,
  JHEP {\bf 1901}, 223 (2019).


\bibitem{Ellis:2003sq} 
  J.~R.~Ellis, M.~Raidal and T.~Yanagida,
  Phys.\ Lett.\ B {\bf 581}, 9 (2004).
  
\bibitem{Giudice:2003jh} 
  G.~F.~Giudice, A.~Notari, M.~Raidal, A.~Riotto and A.~Strumia,
  Nucl.\ Phys.\ B {\bf 685}, 89 (2004).

\bibitem{Peccei:1977hh} 
  R.~D.~Peccei and H.~R.~Quinn,
  Phys.\ Rev.\ Lett.\  {\bf 38}, 1440 (1977).
  
\bibitem{Preskill:1982cy} 
  J.~Preskill, M.~B.~Wise and F.~Wilczek,
  Phys.\ Lett.\  {\bf 120B}, 127 (1983).
  
\bibitem{Abbott:1982af} 
  L.~F.~Abbott and P.~Sikivie,
  Phys.\ Lett.\  {\bf 120B}, 133 (1983).
  
\bibitem{Dine:1982ah} 
  M.~Dine and W.~Fischler,
  Phys.\ Lett.\  {\bf 120B}, 137 (1983).
  
  \bibitem{Sikivie:1982qv} 
  P.~Sikivie,
  Phys.\ Rev.\ Lett.\  {\bf 48}, 1156 (1982).
  
  
\bibitem{Vilenkin:1982ks} 
  A.~Vilenkin and A.~E.~Everett,
  Phys.\ Rev.\ Lett.\  {\bf 48}, 1867 (1982).
  
\bibitem{Kawasaki:2014sqa} 
  M.~Kawasaki, K.~Saikawa and T.~Sekiguchi,
  Phys.\ Rev.\ D {\bf 91}, no. 6, 065014 (2015).
  

\bibitem{globalfit}
NuFIT 5.0, \url{www.nu-fit.org}, 2020.

\bibitem{deSalas:2017kay}
  P.~F.~de Salas, D.~V.~Forero, C.~A.~Ternes, M.~Tortola and J.~W.~F.~Valle,
  Phys.\ Lett.\ B {\bf 782} (2018) 633.

\bibitem{globalfit2}
Valencia-Globalfit, \url{http://globalfit.astroparticles.es/}, 2018.

\bibitem{Buchmuller:2004nz} 
  W.~Buchmuller, P.~Di Bari and M.~Plumacher,
  Annals Phys.\  {\bf 315}, 305 (2005).

\bibitem{Ade:2015xua} 
  P.~A.~R.~Ade {\it et al.} [Planck Collaboration],
  Astron.\ Astrophys.\  {\bf 594}, A13 (2016).
  
\bibitem{Covi:1996wh} 
  L.~Covi, E.~Roulet and F.~Vissani,
  Phys.\ Lett.\ B {\bf 384}, 169 (1996).
  
\bibitem{Abada:2006ea} 
  A.~Abada, S.~Davidson, A.~Ibarra, F.-X.~Josse-Michaux, M.~Losada and A.~Riotto,
  JHEP {\bf 0609}, 010 (2006).
  
\bibitem{Branco:2011zb} 
  G.~C.~Branco, R.~G.~Felipe and F.~R.~Joaquim,
  Rev.\ Mod.\ Phys.\  {\bf 84}, 515 (2012).

\bibitem{Barbieri:1999ma} 
  R.~Barbieri, P.~Creminelli, A.~Strumia and N.~Tetradis,
  Nucl.\ Phys.\ B {\bf 575}, 61 (2000).
 
  
\bibitem{Antusch:2011nz} 
  S.~Antusch, P.~Di Bari, D.~A.~Jones and S.~F.~King,
  Phys.\ Rev.\ D {\bf 86}, 023516 (2012).
 
\bibitem{Casas:2001sr} 
  J.~A.~Casas and A.~Ibarra,
  Nucl.\ Phys.\ B {\bf 618}, 171 (2001).

\end{thebibliography}
\end{document}